\documentclass[pdflatex,sn-mathphys-num]{sn-jnl}
\usepackage{gensymb}

\usepackage{graphicx}%
\usepackage{multirow}%
\usepackage{amsmath,amssymb,amsfonts}%
\usepackage{amsthm}%
\usepackage{mathrsfs}%
\usepackage[title]{appendix}%
\usepackage{xcolor}%
\usepackage{textcomp}%
\usepackage{manyfoot}%
\usepackage{booktabs}%
\usepackage{algorithm}%
\usepackage{algorithmicx}%
\usepackage{algpseudocode}%
\usepackage{listings}%

\usepackage{float}

\theoremstyle{thmstyleone}%
\theoremstyle{thmstyletwo}%

\theoremstyle{thmstylethree}%

\begin{document}

\title[Article Title]{Phase-shifting structured illumination with polarization-encoded metasurface}

\author[1]{\fnm{Linzhi} \sur{Yu}}\email{linzhi.yu@tuni.fi}

\author[1]{\fnm{Jesse} \sur{Pietila}}\email{jesse.pietila@tuni.fi}

\author[1]{\fnm{Haobijam J.} \sur{Singh}}\email{johnson.singh@tuni.fi}

\author*[1,2]{\fnm{Humeyra} \sur{Caglayan}}\email{h.caglayan@tue.nl}

\affil*[1]{\orgdiv{Department of Physics}, \orgname{Tampere University}, \city{Tampere}, \postcode{33720}, \country{Finland}}

\affil[2]{\orgdiv{Department of Electrical Engineering and Eindhoven Hendrik Casimir Institute}, \orgname{Eindhoven University of Technology}, \city{Eindhoven}, \postcode{5600 MB}, \country{The Netherlands}}

\abstract{Phase-shifting structured illumination is a powerful technique used across diverse imaging modalities, including 3D surface measurement, quantitative phase imaging, and super-resolution microscopy. However, conventional implementations often rely on mechanically driven or optoelectronically complex systems, limiting their compactness, stability, and integration. Here, we present a polarization-controlled dielectric metasurface that generates phase-shifting fringe patterns in the visible spectrum, enabling compact and robust structured light projection. The metasurface encodes distinct phase gratings for orthogonal polarizations, producing fringe patterns with relative lateral displacements that vary according to the polarization of the transmitted light. We experimentally demonstrate high-quality fringe generation and apply the structured illumination in a fringe projection profilometry system for 3D surface measurement of different objects. The metasurface integrates multiple phase-shifting steps into a single static device, offering a millimeter-scale footprint and compatibility with polarization multiplexing. This approach introduces a compact, passive solution for structured light generation with broad potential in next-generation optical metrology and advanced computational imaging.}

\maketitle

\section{Introduction}\label{sec1}

Phase-shifting structured illumination plays a central role in a wide range of optical imaging applications~\cite{saxena2015structured}, including three-dimensional (3D) surface measurement~\cite{geng2011structured,zhang2018high}, quantitative phase imaging~\cite{zhang1998three,bhaduri2014diffraction}, and super-resolution microscopy~\cite{wu2018faster,gustafsson2000surpassing}. In these modalities, sinusoidal fringe patterns with controlled phase offsets are projected onto objects, and the resulting intensity maps are processed to reconstruct surface topography, optical path length variations, or fine spatial features beyond the diffraction limit. A key requirement for such systems is the ability to generate high-quality fringe patterns with tunable phase shifts. Traditional methods achieve phase shifting by mechanically moving gratings~\cite{mauvoisin1994three,schreiber1997lateral} or adjusting the relative optical path length between interfering beams~\cite{yoneyama2003three,lai2007surface}. While effective, these approaches suffer from limited stability, slow response times, and poor scalability. More recently, active spatial modulation devices such as spatial light modulators~\cite{coggrave1999high,li2015extended}, digital micromirror devices~\cite{dan2013dmd,van2016real}, and acousto-optic deflectors~\cite{dupont2010structured,guan2014dynamic} have been introduced to enhance programmability and speed. However, their relatively large pixel sizes and bulky optical setups hinder their integration into compact, high-resolution systems.

Metasurfaces offer a promising path forward. These ultrathin optical elements, composed of subwavelength-spaced meta-atoms, enable spatially varying modulation of light’s amplitude, phase, and polarization in a single, nanostructured layer~\cite{yu2011light,yu2014flat}. Plasmonic metasurfaces have demonstrated phase-shifting structured light via surface plasmon interference~\cite{wang2012subwavelength,wei2014wide,zhang2018polarization,tan2020polarization}, but suffer from high optical loss—particularly in the visible range~\cite{meinzer2014plasmonic}. Based on high-index, low-loss materials, dielectric metasurfaces have emerged as a superior alternative~\cite{lin2014dielectric}. Yet, existing designs typically generate fixed patterns~\cite{jing2023active} or require complex co-modulation of both the illumination and detection pathways~\cite{zhou2024far}, limiting their flexibility and complicating integration into compact or adaptive imaging systems.

In this work, we present a polarization-controlled dielectric metasurface capable of generating high-quality phase-shifting structured illumination in the visible spectrum. The design leverages birefringent titanium dioxide (TiO\textsubscript{2}) nanopillars to encode two distinct lateral phase profiles for orthogonal polarization states. By adjusting the polarization direction of the transmitted light, multiple fringe patterns with controlled phase offsets can be selectively accessed. This enables phase-shifting functionality to be embedded into a single, static, and ultra-compact optical element. Compared to existing approaches, this method offers high fringe fidelity and system stability while reducing the optical complexity of the system.
Furthermore, we implemented this approach in a fringe projection profilometry system and demonstrated robust 3D surface reconstruction with a polarization-multiplexed three-step phase-shifting scheme.  Beyond 3D surface measurement, the polarization-dependent fringe encoding strategy shown here can be extended to structured illumination microscopy~\cite{wu2018faster}, quantitative phase imaging~\cite{huang2024quantitative}, and digital holography~\cite{yamaguchi2006phase}, where compact and reconfigurable multi-pattern illumination is essential. The proposed platform offers significant potential for a wide range of optical systems from biomedical imaging~\cite{webb2022introduction} and machine vision~\cite{steger2018machine} to precision metrology~\cite{shimizu2021insight}.

\section{Principle}\label{sec2}

\subsection{Principle of polarization-controlled phase-shifting structured illumination}\label{subsec2_1}

\begin{figure}[H]
\centering
\includegraphics[width=0.5\textwidth]{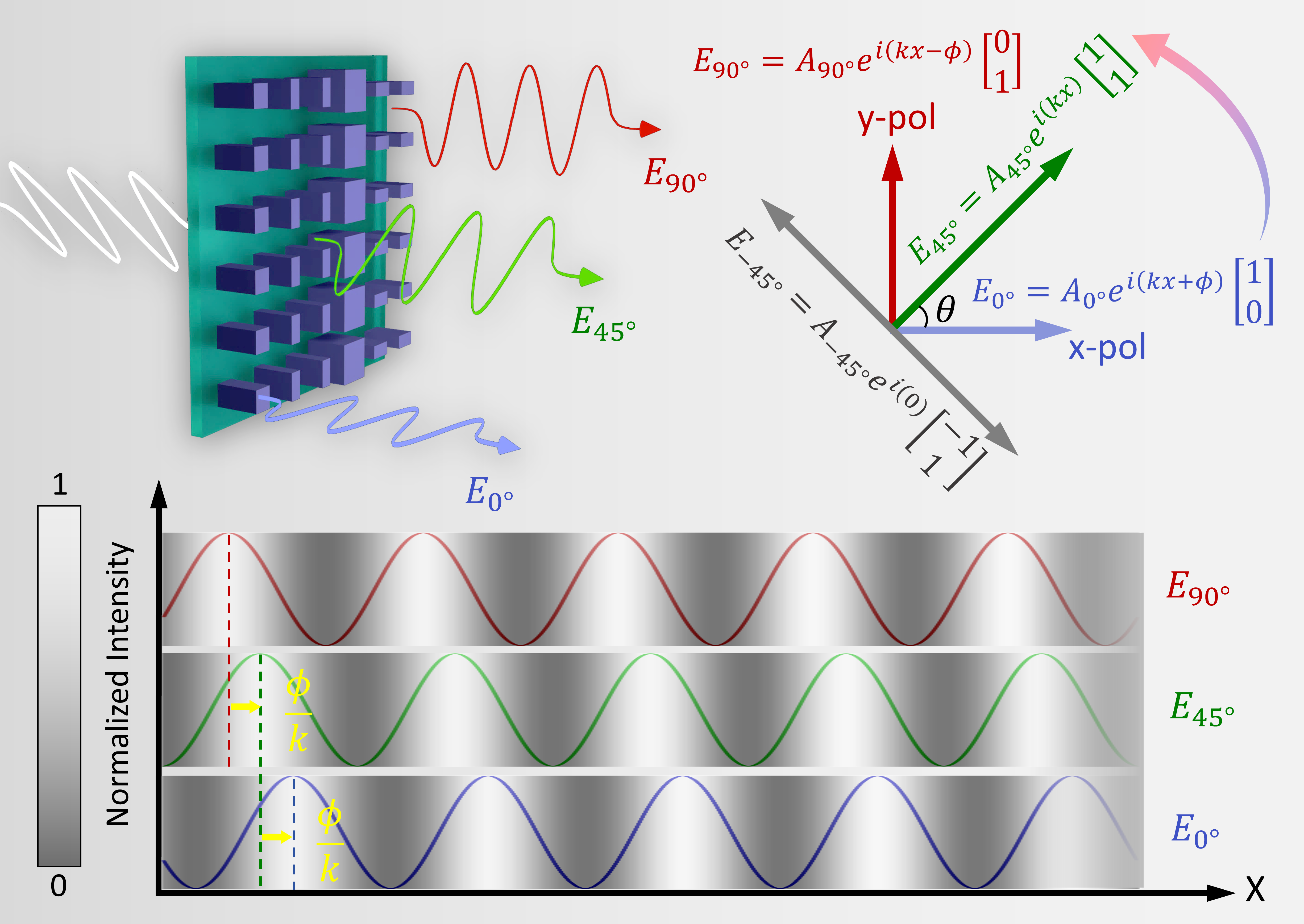}
\caption{\textbf{Schematic illustration of polarization-controlled phase-shifting structured illumination using a metasurface.} The metasurface produces distinct fringe patterns with controlled lateral phase shifts depending on the polarization state of the transmitted beam. This enables compact, multi-pattern structured light generation through polarization control.}\label{fig 1}
\end{figure}

The principle of polarization-controlled phase-shifting structured illumination is illustrated in Figure~\ref{fig 1}. The metasurface is designed to encode two phase grating profiles for x- and y-polarized light, which are structurally identical but laterally offset. This leads to polarization-dependent fringe patterns whose lateral phase shift can be tuned by adjusting the polarization of the transmitted light. For x-polarized light (\(\theta = 0\degree\)), the transmitted electric field can be expressed as:
\begin{equation}
E_{0\degree}(x) = A \cos(kx + \phi),
\end{equation}
yielding the intensity distribution:
\begin{equation}
I_{0\degree}(x) = \frac{A^2}{2} \left(1 + \cos(2kx + 2\phi)\right).
\end{equation}
Similarly, for y-polarized light (\(\theta = 90\degree\)):
\begin{equation}
E_{90\degree}(x) = A \cos(kx - \phi), \quad
I_{90\degree}(x) = \frac{A^2}{2} \left(1 + \cos(2kx - 2\phi)\right).
\end{equation}
For an arbitrary linear polarization angle \(\theta\), the resulting field is a coherent superposition of the x- and y-polarized components, leading to an intensity profile:
\begin{equation}
I_\theta(x) = A^2 \left[ \cos^2(\theta) \cos^2(kx + \phi) + \sin^2(\theta) \cos^2(kx - \phi) \right].
\end{equation}
In particular, for \(\theta = 45\degree\), where both polarization components contribute equally, the transmitted field simplifies to:
\begin{equation}
E_{45\degree}(x) = \sqrt{2}A \cos(\phi) \cos(kx), \quad
I_{45\degree}(x) = A^2 \cos^2(\phi) \left(1 + \cos(2kx)\right).
\end{equation}
This analysis shows that the polarization state of the light governs the lateral phase shift of the projected sinusoidal fringe pattern. In this work, we use polarization angles of \(0\degree\), \(45\degree\), and \(90\degree\) to generate three fringe patterns with relative phase offsets of \(-\frac{2\pi}{3}\), \(0\), and \(+\frac{2\pi}{3}\), respectively. These patterns form a polarization-controlled phase-shifting sequence suitable for standard three-step phase retrieval. While a three-step configuration is demonstrated here, the concept is readily extendable to multi-step phase-shifting schemes by appropriately selecting polarization states. The detailed derivation of the polarization-dependent intensity profiles is provided in Supplementary Information Section 1. This polarization-encoded phase modulation principle underlies the metasurface design presented in this work.

\subsection{Metasurface design}\label{subsec2_2}

\begin{figure}[H]
\centering
\includegraphics[width=1\textwidth]{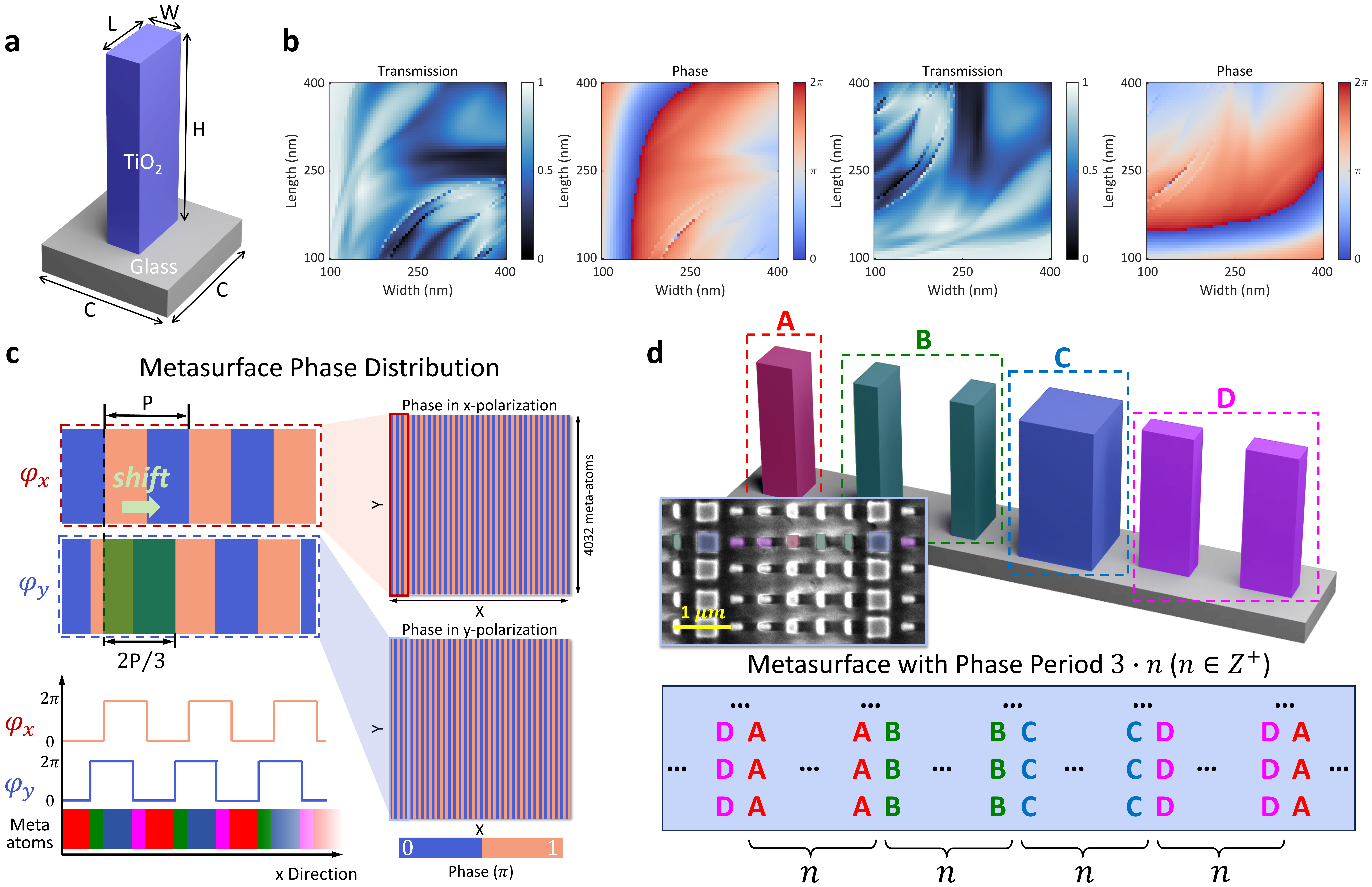}
\caption{\textbf{Metasurface design.} \textbf{a} Schematic of TiO\textsubscript{2} rectangular nanopillar meta-atoms forming the metasurface. \textbf{b} Simulated transmission and phase responses of the meta-atoms under x- and y-polarized illumination, as functions of varying pillar width and length. \textbf{c} Polarization-dependent phase maps encoded into the metasurface, showing distinct grating profiles for x- and y-polarized incidence. \textbf{d} Structural layout of the metasurface and scanning electron microscope image of a fabricated sample with \(n=1\).}\label{fig 2}
\end{figure}

To implement polarization-dependent phase modulation for structured light projection, we designed a dielectric metasurface composed of rectangular TiO\textsubscript{2} nanopillars fabricated on a glass substrate, as shown in Figure~\ref{fig 2}a. Each nanopillar has a fixed height of 600\,nm and is characterized by its width \(W\) and length \(L\), independently influencing the phase delays imparted to x- and y-polarized light. The meta-atoms are arranged in a square lattice with a subwavelength period of 450\,nm, enabling high-efficiency modulation. To guide the metasurface design, we numerically simulated the polarization-resolved transmission and phase responses of individual meta-atoms using finite element analysis at the operational wavelength of 532\,nm. The resulting design library, shown in Figure~\ref{fig 2}b, maps the achievable phase shifts and transmission efficiencies as functions of the meta-atom dimensions under both x- and y-polarized illumination. This library serves as the basis for assigning geometries to implement polarization-dependent phase profiles across the metasurface.

Based on the simulated meta-atom response library, we designed a metasurface that simultaneously encodes two binary phase gratings for x- and y-polarized light, laterally offset by \(2P/3\), where \(P\) is the grating period (Figure~\ref{fig 2}c)~\cite{su1992automated,zuo2018phase}. These gratings generate sinusoidal fringe patterns with identical intensity profiles but lateral displacements that depend on the polarization state. The design is composed of six meta-atoms that implement the required discrete phase steps. Among them, only four have distinct geometries; the remaining two are structural duplicates used to complete the phase sequence. For clarity, these are grouped into four categories labeled A, B, C, and D. As illustrated in Figure~\ref{fig 2}d, the six-element phase unit is periodically repeated \(n\) times across the metasurface, yielding a total grating period of \(3n\) pixels. In our implementation, we set \(n = 20\), resulting in a total period of 60 pixels. Detailed optical responses of the four structural types are provided in Supplementary Information Section 4.

The metasurface functions as a polarization-multiplexed diffractive element that modulates orthogonal polarization components independently. Under illumination by a linearly polarized plane wave with equal x- and y-polarized components, the encoded phase gratings generate a composite transmitted field comprising multiple spatial frequency components. At the metasurface plane (\(z = 0\)), the field can be described by a Fourier series:
\begin{equation}
E_{\text{trans}}(x, 0) = \sum_{m} A_m e^{i (k + mK)x},
\end{equation}
where \(k\) is the incident wavevector, \(K = 2\pi/P\) is the grating spatial frequency, and \(A_m\) are Fourier coefficients determined by the binary phase profile. Upon propagation, each diffraction order accumulates an additional phase:
\begin{equation}
E(x, z) = \sum_{m} A_m e^{i (k + mK)x} e^{i \sqrt{k^2 - (k + mK)^2} z}.
\end{equation}
Higher-order components with \(|mK| > k\) are evanescent or strongly diverging, and are effectively suppressed. At a finite distance, the observed field primarily consists of the fundamental modes:
\begin{equation}
E_{\text{obs}}(x, z) \approx A_0 e^{ikx} + A_1 e^{i(k + K)x} + A_{-1} e^{i(k - K)x},
\end{equation}
which corresponds to a spatially modulated wave:
\begin{equation}
E_{\text{obs}}(x, z) \approx A \cos(Kx + \phi),
\end{equation}
where \(A\) and \(\phi\) depend on the amplitudes and relative phases of the retained components. The propagation behavior confirms that binary phase gratings, when appropriately designed, evolve into smooth sinusoidal patterns suitable for structured illumination. Crucially, the lateral phase shift between the polarization-encoded gratings is preserved in the transmitted field, providing a robust mechanism for polarization-controlled fringe generation. This wavefront shaping capability underpins the phase-shifting illumination demonstrated in the experiments that follow. To implement this design, the metasurface was fabricated using standard electron-beam lithography followed by dry etching, as detailed in the Methods section and Supplementary Information Section 5.

\section{Experiment}\label{sec3}

\subsection{Characterization of polarization-controlled phase-shifting fringes}\label{subsec3_1}

\begin{figure}[H]
\centering
\includegraphics[width=1\textwidth]{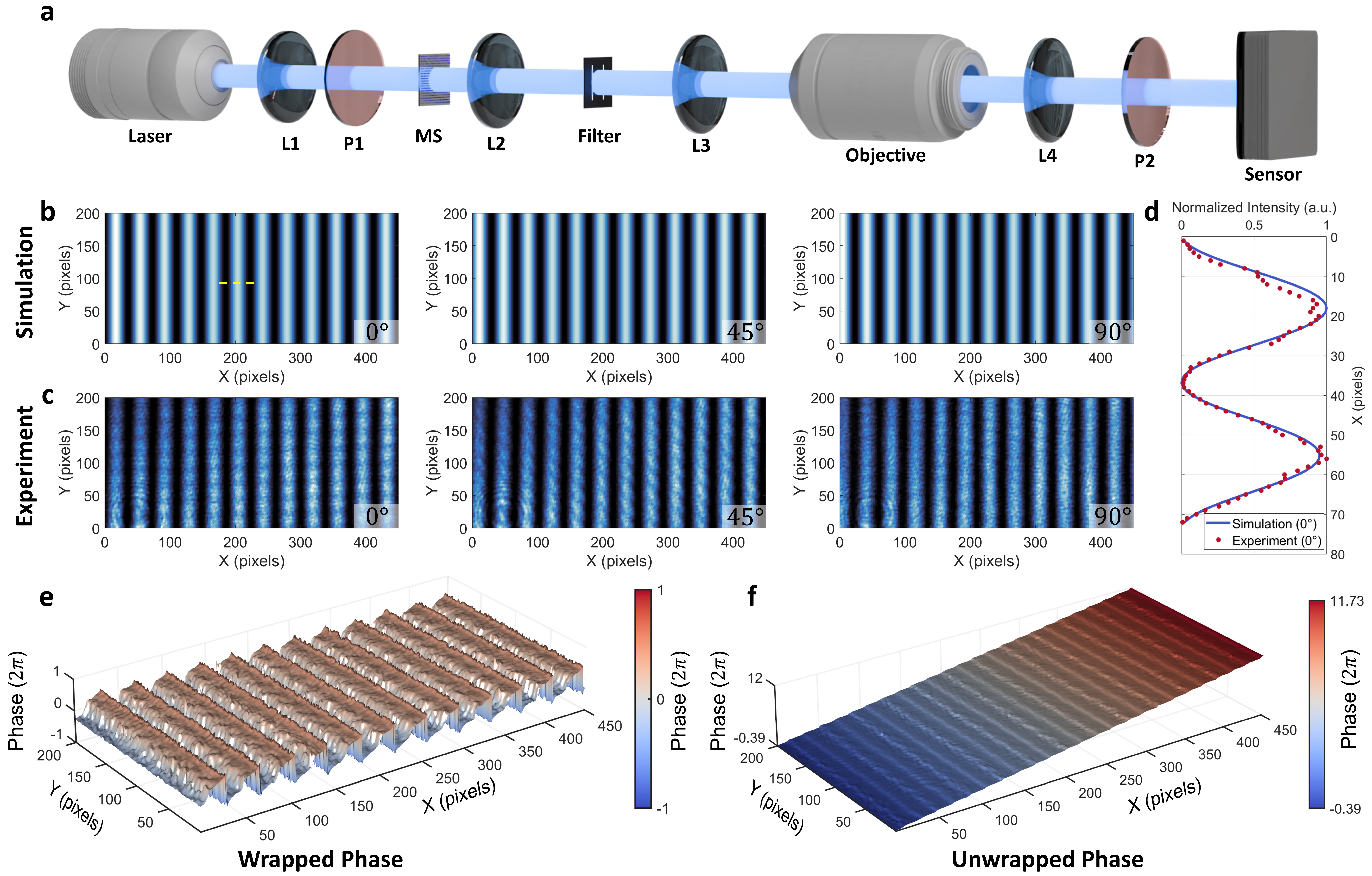}
\caption{\textbf{Characterization of polarization-controlled phase-shifting fringes.} \textbf{a} Experimental setup for fringe pattern generation and analysis. L, lens; P, polarizers; MS, metasurface. \textbf{b} Simulated fringe patterns generated by the metasurface under input polarization angles of 0\textdegree{}, 45\textdegree{}, and 90\textdegree{}. \textbf{c} Corresponding experimentally captured fringe patterns. \textbf{d} Comparison of simulated and experimental lateral intensity profiles. \textbf{e} Wrapped phase distribution retrieved from the patterns in \textbf{c} using a three-step phase-shifting algorithm. \textbf{f} Unwrapped phase distribution from \textbf{e}, showing a continuous phase profile.}\label{fig 3}
\end{figure}

To evaluate the performance of the polarization-controlled metasurface, we experimentally characterized the quality of the generated phase-shifting structured illumination under different transmitted polarization states. This assessment focused on both the spatial fidelity of the projected fringe patterns and the accuracy of the retrieved phase information. The experimental setup is illustrated in Figure~\ref{fig 3}a. A collimated, linearly polarized laser beam with equal x- and y-polarized components was directed onto the metasurface using a polarizer (P1). The resulting polarization-dependent phase-shifting fringe patterns were recorded at transmitted polarization angles of 0\textdegree{}, 45\textdegree{}, and 90\textdegree{}, selected via an analyzer (P2). Figure~\ref{fig 3}b shows the simulated fringe patterns corresponding to the three polarization angles, while Figure~\ref{fig 3}c presents the experimentally captured results. The lateral intensity profiles extracted from the experimental image closely match the simulated prediction (Figure~\ref{fig 3}d), confirming the fidelity of the metasurface in reproducing the designed fringe patterns. To retrieve the encoded phase information, we applied a standard three-step phase-shifting algorithm to the recorded fringe patterns. Given three phase-shifting fringe patterns \(I_{0\degree}\), \(I_{45\degree}\), and \(I_{90\degree}\), the wrapped phase \(\phi(x)\) is calculated as:
\begin{equation}
\phi(x) = \tan^{-1} \left( \frac{\sqrt{3} \, (I_{0\degree} - I_{90\degree})}{2I_{45\degree} - I_{0\degree} - I_{90\degree}} \right).
\end{equation}
The resulting wrapped phase distribution is shown in Figure~\ref{fig 3}e. A spatial phase unwrapping algorithm~\cite{su2010phase} was then applied to recover a continuous phase map, as shown in Figure~\ref{fig 3}f. The smooth and well-defined phase profiles obtained from the measurements demonstrate the high optical quality and stability of the structured illumination produced by the metasurface, validating its effectiveness for precise and compact phase-shifting light projection.

\subsection{3D profilometry with metasurface-generated phase-shifting fringes}\label{subsec3_2}

\begin{figure}
\centering
\includegraphics[width=1\textwidth]{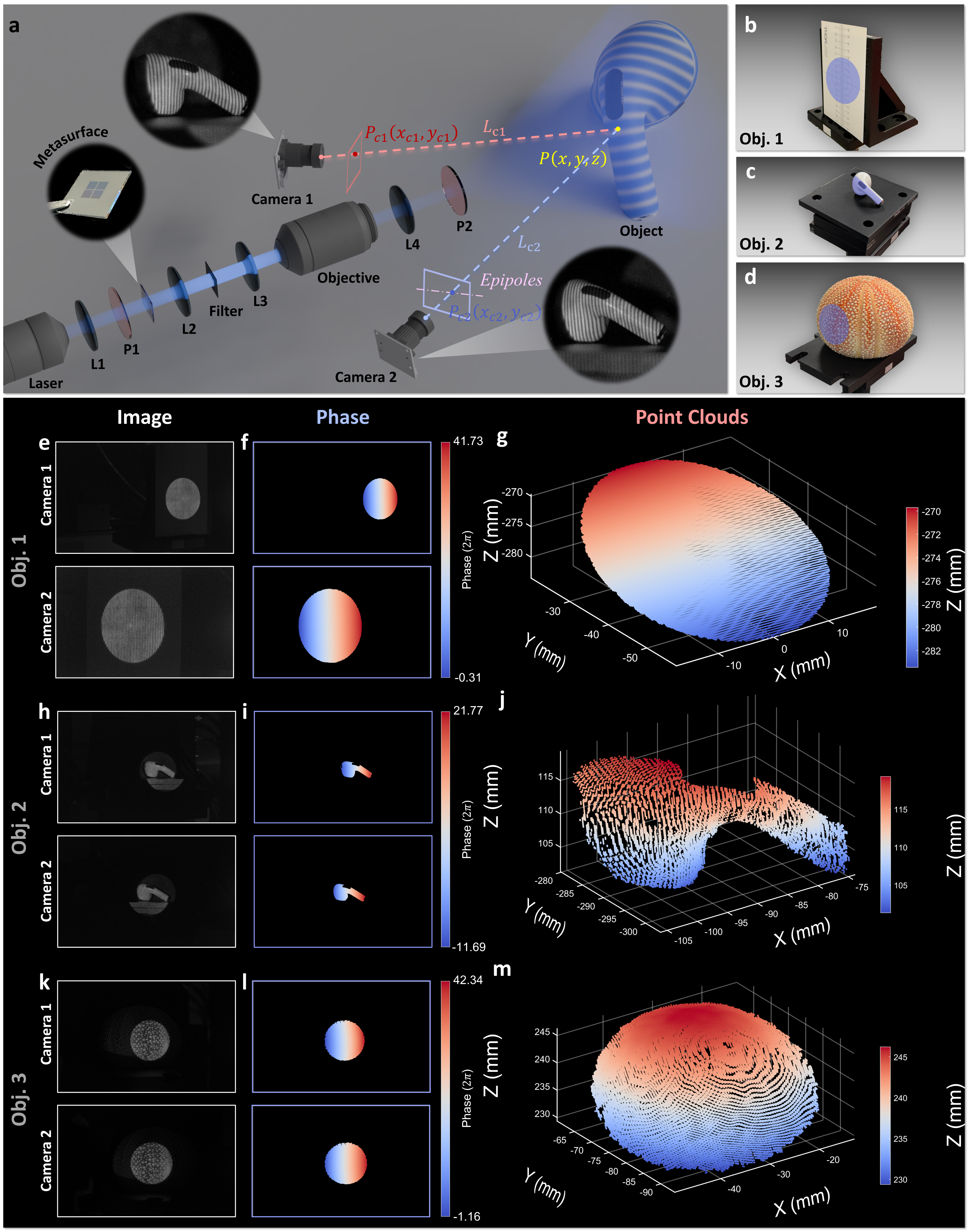}
\caption{\textbf{Fringe projection profilometry using metasurface-generated phase-shifting structured illumination.} \textbf{a} Schematic illustration of the working principle and experimental setup for 3D surface measurement. \textbf{b}, \textbf{c}, \textbf{d} Photographs of the test samples: a business card, an earphone, and a sea urchin shell. Shaded regions indicate the selected measurement areas. \textbf{e}, \textbf{h}, \textbf{k} Captured structured illumination patterns at 0\textdegree{} transmitted polarization for the respective samples. \textbf{f}, \textbf{i}, \textbf{l} Respective retrieved phase distributions. \textbf{g}, \textbf{j}, \textbf{m} Respective reconstructed 3D point clouds.}\label{fig 4}
\end{figure}

To demonstrate the practical applicability of the metasurface-generated phase-shifting structured illumination, we integrated it into a fringe projection profilometry system for 3D surface measurement. The working principle and experimental setup are shown in Figure~\ref{fig 4}a. Polarization-controlled phase-shifting fringe patterns were projected onto the object surface and captured simultaneously by two cameras arranged in a calibrated stereo configuration.

The stereo system was calibrated using Zhang’s method~\cite{zhang2002flexible}, which estimates the intrinsic parameters of each camera as well as the relative rotation and translation between them~\cite{bradski2008learning}. At each viewpoint, the local phase distribution was retrieved using a three-step phase-shifting algorithm~\cite{zuo2018phase,OpticalShopTesting}, and these phase values served as discriminative descriptors for establishing pixel correspondences across the stereo views. Stereo rectification was applied to simplify the epipolar geometry, such that corresponding pixels lie along the same horizontal scanline~\cite{bradski2008learning}. Let
\begin{equation}
P_{C1} = (u_1, v), \quad P_{C2} = (u_2, v)
\end{equation}
denote the image coordinates of a 3D point \(P\) in \textit{Camera 1} and \textit{Camera 2}, respectively. The horizontal disparity is defined as
\begin{equation}
d = u_1 - u_2.
\end{equation}
The perspective projection of a 3D point \(P = (x, y, z, 1)^\top\) into image coordinates \(\mathbf{x}_i = (u_i, v_i, 1)^\top\) in camera \(i\) follows:
\begin{equation}
\lambda_i\,\mathbf{x}_i = \mathbf{A}_i[\mathbf{R}_i \,\, \mathbf{t}_i]\,\mathbf{P}, \quad i \in \{1, 2\},
\end{equation}
where \(\mathbf{A}_i\) is the intrinsic matrix, and \(\mathbf{R}_i\), \(\mathbf{t}_i\) are the extrinsic parameters of camera \(i\). The scalar \(\lambda_i\) represents a projective scale factor. The relative pose between the two cameras is given by:
\begin{equation}
\mathbf{R}_{\mathrm{rel}} = \mathbf{R}_{C2}\mathbf{R}_{C1}^\top, \quad
\mathbf{t}_{\mathrm{rel}} = \mathbf{t}_{C2} - \mathbf{R}_{\mathrm{rel}}\,\mathbf{t}_{C1}.
\end{equation}
Given a stereo pixel pair \((\mathbf{x}_1, \mathbf{x}_2)\) matched via phase similarity, the corresponding 3D point is recovered by triangulation. Ideally, the viewing rays from both cameras intersect at a single point in space. In practice, due to calibration uncertainties and noise, the 3D coordinates are computed by minimizing the squared distance to both rays:
\begin{equation}
\min_{P} \left( \|P - L_{C1}\|^2 + \|P - L_{C2}\|^2 \right),
\end{equation}
where \(L_{C1}\) and \(L_{C2}\) represent the lines extending from the respective camera centers through the image points~\cite{hartley2004multiple,salvi2010state}. Additional details on the stereo calibration and 3D reconstruction process are provided in Supplementary Information Sections 2 and 3.

Three representative objects with diverse geometries and surface properties—a business card, an earphone, and a sea urchin shell—were selected as test samples (Figures~\ref{fig 4}b–d). Polarization-controlled phase-shifting fringe patterns were projected onto each object and captured from two viewpoints using synchronized stereo cameras (Figures~\ref{fig 4}e, h, k). To enhance the quality of the patterns, the captured images were processed by the frequency domain filtering to suppress noise~\cite{gonzalez2006}. Phase distributions were then computed using a three-step phase-shifting algorithm (Figures~\ref{fig 4}f, i, l), which served as the basis for establishing pixel correspondences across the stereo views. The region of interest for each object was extracted using a mask generated from the local modulation intensity of the structured light, ensuring that only areas with sufficient fringe contrast contributed to the 3D reconstruction~\cite{neil1997method}. 

The reconstructed 3D point clouds (Figures~\ref{fig 4}g, j, m) clearly capture the surface morphology and fine structural features of the test samples, demonstrating reliable performance across varying object geometries and material properties. These results validate the metasurface’s capability to support robust fringe projection for high-quality 3D surface measurement. By encoding polarization-dependent phase shifts into a static, compact optic, the system minimizes optical complexity and footprint, offering a promising alternative to bulkier solutions that rely on dynamic spatial light modulation.

\section{Conclusion}\label{sec4}

In conclusion, we have demonstrated a compact dielectric metasurface capable of generating polarization-controlled phase-shifting structured illumination. By leveraging birefringent TiO\textsubscript{2} nanopillars to encode distinct phase profiles for orthogonal polarizations, the metasurface enables tunable fringe pattern generation through simple polarization control. This static optical component integrates multiple phase-shifting states into a single planar layer, without requiring active modulation or tunable elements, offering a robust and compact alternative to conventional structured light systems. We experimentally validated the metasurface in a fringe projection profilometry setup, achieving accurate 3D surface reconstruction of complex samples with diverse geometries and materials. The results highlight the metasurface’s capability to serve as a high-quality fringe pattern generator in a simplified optical architecture, dramatically reducing system size and alignment complexity. Beyond 3D profilometry, the principle of polarization-controlled phase modulation offers a versatile foundation for structured illumination in a variety of computational imaging techniques. Its compact, passive configuration and polarization-based phase control offer a practical route to multi-state structured illumination in miniaturized imaging and sensing platforms. By integrating structured illumination control into a metasurface platform, this work lays the groundwork for future photonic components that bridge meta-optics and computational imaging in biomedical, industrial, and scientific domains.

\section{Methods}\label{sec5}

\subsection{Numerical simulation and results analysis}\label{subsec5_1} The optical responses of TiO\textsubscript{2} nanopillars with varying dimensions, as shown in Figure~\ref{fig 2}b, were simulated using the finite-difference frequency-domain method implemented in the RF module of COMSOL Multiphysics 6.2. Each unit cell consisted of a rectangular TiO\textsubscript{2} nanopillar placed on a float glass substrate. Periodic boundary conditions were applied along the lateral directions, and perfectly matched layers were used in the light incidence direction to absorb outgoing waves. The refractive index of TiO\textsubscript{2} was measured by ellipsometry and found to be 2.47 with negligible absorption at the design wavelength of 532 nm. The refractive index of the glass substrate was set to 1.53. Numerical results in Figure~\ref{fig 3}b and data analysis of the experimental results in Figures~\ref{fig 3}c–f and \ref{fig 4}e–m were performed using MATLAB R2022b.

\subsection{Metasurface fabrication}\label{subsec5_2} The metasurface was fabricated by depositing a 600 nm thick TiO\textsubscript{2} film onto glass substrates using ion beam sputtering. A layer of polymethyl methacrylate (PMMA 950K A4) electron-beam resist was then spin-coated and patterned using electron-beam lithography with a pre-designed 2D layout. Following development in a methyl isobutyl ketone:isopropyl alcohol (MIBK:IPA, 1:3) solution, a chromium (Cr) hard mask was deposited by electron-beam evaporation and patterned via a lift-off process. The underlying TiO\textsubscript{2} layer was etched using reactive ion etching to transfer the pattern, and the residual Cr mask was subsequently removed using a chromium etchant. The final metasurface structure was thus revealed on the substrate. Additional fabrication details are provided in Supplementary Information Section 5.

\subsection{Experimental setup}\label{subsec5_3} 
The experimental setups used in this study are shown in Figure~\ref{fig 3}a and Figure~\ref{fig 4}a, differing only in the choice of tube lens (L4) to accommodate different magnifications. A collimated 532 nm laser beam from a Raman laser source (WiTec, Oxford Instruments) was expanded and collimated using an achromatic lens (L1, AC254-050-A-ML, Thorlabs), followed by a linear polarizer (P1, LPVISE100-A, Thorlabs) to ensure equal amplitude components at 0\textdegree{} and 90\textdegree{} polarizations. The beam was then modulated by the metasurface. A 4\(f\) system (L2, L3: AC254-075-AB-ML, Thorlabs) with a spatial filter at the Fourier plane was used to simulate free-space propagation in a compact setup by suppressing unwanted diffraction orders, thereby enabling the formation of the expected structured illumination pattern. The modulated beam was subsequently magnified by an objective lens (EC Epiplan-Neofluar 20×/0.5, ZEISS) and tube lens (L4). A second polarizer (P2, LPVISE100-A, Thorlabs), placed after the tube lens, served as an analyzer to select the desired polarization-dependent phase shift. For the evaluation of phase-shifting pattern quality (Figure~\ref{fig 3}), the tube lens focal length was set to 150 mm (AC254-150-AB-ML, Thorlabs), and images were captured using a CMOS sensor (ORCA-Fusion C14440-20UP, Hamamatsu). For the fringe projection profilometry experiments (Figure~\ref{fig 4}), the tube lens focal length was set to 200 mm (AC254-200-AB-ML, Thorlabs). The illuminated objects were imaged by two cameras (Camera 1, Camera 2: acA1920-50gm, Basler), each equipped with a 16 mm focal length lens. The test samples included a business card (Thorlabs), Apple AirPods Pro 2 (Apple Inc.), and a sea urchin shell collected from the Lofoten islands, Norway.

\backmatter

\bmhead{Supplementary Information}
Supplementary information is available.

\bmhead{Acknowledgements}
L.Y. and H.C. acknowledge financial support from the European Union’s Horizon 2020 research and innovation programme under the Marie Skłodowska-Curie grant agreement No 956770.

\bmhead{Conﬂict of Interest}
The authors declare no conﬂict of interest.

\bmhead{Data Availability Statement}
The data that support the ﬁndings of this study are available from the cor-responding author upon reasonable request.

\bmhead{Keywords}
metasurfaces,phase-shifting structured illumination,fringe projection profilometry,3D measurement.

\bibliography{sn-bibliography}

\end{document}